\documentstyle[12pt,bezier]{article}

\begin{document}
\pagestyle{plain}
\setcounter{page}{1}
\baselineskip25pt

\renewcommand{\baselinestretch}{2}

\newcommand{\be}{\begin{eqnarray}}
\newcommand{\ee}{\end{eqnarray}}

\newcommand{\disk}{{\rm D}}
\newcommand{\grad}{\nabla}
\newcommand{\rightvbar}[1]{\left. #1 \right|}
\newcommand{\tr}{\mathop{\rm tr}}
\newcommand{\sym}{\mathop{\rm sym}}
\newcommand{\id}{1}

\newcommand{\sA}{{\cal A}}
\newcommand{\sB}{{\cal B}}
\newcommand{\sD}{{\cal D}}
\newcommand{\sL}{{\cal L}}
\newcommand{\sM}{{\cal M}}
\newcommand{\sO}{{\cal O}}

\newcommand{\va}{\vec{a}}
\newcommand{\vb}{\vec{b}}
\newcommand{\vn}{\vec{n}}
\newcommand{\vv}{\vec{v}}
\newcommand{\vw}{\vec{w}}
\newcommand{\vx}{\vec{x}}
\newcommand{\vJ}{\vec{J}}
\newcommand{\valpha}{\vec{\alpha}}
\newcommand{\dalpha}{\dot{\valpha}}
\newcommand{\vtau}{\vec{\tau}}

\newcommand{\bc}{\bar{c}}
\newcommand{\bs}{\bar{s}}
\newcommand{\bomega}{\bar{\omega}}

\newcommand{\unitr}{\widehat{r}}
\newcommand{\unittheta}{\widehat{\theta}}
\newcommand{\unitphi}{\widehat{\phi}}

\newcommand{\dr}{{\rm\bf d}r}
\newcommand{\dtheta}{{\rm\bf d}\theta}
\newcommand{\dphi}{{\rm\bf d}\phi}
\newcommand{\dt}{{\rm\bf d}t}

\renewcommand{\baselinestretch}{1.0}
\renewcommand{\theequation}{\thesection.\arabic{equation}}
\renewcommand{\vec}[1]{{\bf #1}}
\newcommand{\np}[3]{ {\it Nucl.\ Phys.\ } {\bf B{#1}} ({#2}) {#3}}
\newcommand{\prd}[3]{ {\it  Phys.\ Rev.\ } {\bf D{#1}} ({#2}) {#3}}
\newcommand{\prl}[3]{ {\it  Phys.\ Rev.\ Lett.\ } {\bf {#1}} ({#2}) {#3}}
\newcommand{\zp}[3]{ {\it Z. Phys.\ } {\bf C{#1}} ({#2}) {#3}}
\newcommand{\pl}[3]{ {\it Phys.\ Lett.\ }{\bf B{#1}} ({#2}) {#3}}

\begin{titlepage}

\begin{flushright}
PUPT-1577\\
\vskip-0.4truecm
CTP-2487\\
\vskip-0.4truecm
hep-th/9511104
\end{flushright}
\vspace{10 mm}

\begin{center}
{\huge Screening vs. Confinement in 1+1 Dimensions}

\vspace{5mm}

\end{center}

\vspace{7 mm}

\begin{center}
{\large David J.\ Gross, Igor R.\ Klebanov\\
{\small \it Joseph Henry Laboratories,
Princeton University,
Princeton, New Jersey 08544, USA}  }
\end{center}
\begin{center}
{\large Andrei V.\ Matytsin\\
{\small \it Center for Theoretical Physics, MIT, Cambridge, MA 02139, USA}\\
        and\\
Andrei V. Smilga \\
{\small\it ITEP, B.Cheremushkinskaya 25, Moscow 117259, Russia}\\}

\end{center}

\vspace{5mm}

\renewcommand{\baselinestretch}{1.5} 
\parindent=20pt

\begin{center}
{\bf Abstract}\\
\end{center}
We show that, in 1+1 dimensional gauge theories,
a heavy probe charge is screened by
dynamical massless fermions both in the case when the source and the dynamical
fermions belong to the same representation of the gauge group and,
unexpectedly, in the case when the representation of the probe charge is
smaller than the representation of the massless fermions . Thus, a
fractionally charged  heavy probe is screened by dynamical fermions of integer
charge in the massless Schwinger model, and a colored probe  in the
fundamental representation is screened in $QCD_2$ with adjoint massless
Majorana fermions. The screening disappears and confinement is
restored as soon as the dynamical
fermions are given a non-zero mass. For small masses,
the string tension is given by the product of the
light fermion mass and the fermion
condensate with a known numerical coefficient.

Parallels with 3+1 dimensional $QCD$ and supersymmetric gauge theories are
discussed.

\vfill
\end{titlepage}

\section{Introduction}
The proof of confinement in $QCD$ remains a major unsolved
problem. The heuristic picture of confinement is well known.
In pure Yang-
Mills theory, the static potential between heavy probe charges in
the fundamental color representation is believed to grow linearly at large
distances
  \be
  \label{VQQ}
V_{Q \bar Q}(r) \sim \sigma r .
  \ee
This corresponds to the famous area law behavior of the Wilson loop vacuum
expectation value,
  \be
  \label{area}
 <W(C)> = \left< \frac 1{N_c} {\rm Tr}P \exp \left\{ig \int_C \hat{A}_\mu
dx_\mu
\right\} \right>\
\sim \ \exp\{-\sigma {\cal A}_C\},
  \ee
for large smooth quasi-planar loops where ${\cal A}_C$ is the area of the
minimal surface with boundary $C$.

It is also well-known that
the area law (\ref{area}) does {\it not} hold in
$QCD$ with dynamical
quarks. The string may tear by creating  a light
quark-antiquark pair so that the color charge of heavy sources is screened
by the dynamical quarks. The potential $V_{Q \bar Q} (r)$
should then  approach a
constant for large $r$, and the Wilson loop average should display a perimeter
law. The same is true for the Wilson loop in
the {\it adjoint} representation in
pure Yang-Mills theory.
  Although the  spectrum  of $ QCD$ with
dynamical quarks contains only colorless states, it is important to distinguish
this {\it screening} picture
from true {\it confinement} with an area law for the Wilson
loop.

Recently, the first $4D$ field theory example where confinement
is proved (at least at a physical level of rigorousness) has
been constructed \cite{Seiberg} . The theory is an ${\cal N} =
2$ supersymmetric $SU(2)$
Yang-Mills theory with an extra term that breaks
${\cal N} = 2$ supersymmetry to ${\cal N} = 1$  giving a
small mass to one of the two adjoint Majorana fermions and its ${\cal N} = 1$
scalar superpartner. Confinement is absent when $m = 0$ but {\it does}
appear for any non-zero $m$. Due to the  special nature of this model,
the confinement affects only the $U(1)$ subgroup of $SU(2)$; the rest
of the group is in the Higgs phase.

In this paper we show
that a similar onset of confinement as a mass is introduced
takes place in some simple 2D
non-abelian models such as the $SU(N)$ gauge theory
coupled to Majorana fermions in the adjoint representation,
when the heavy probe
charges are in the fundamental representation.
Here the entire $SU(N)$ is in a screening phase
for  vanishing fermion mass but becomes confining as the mass
is turned on. An abelian prototype of this phenomenon,
the transition from screening to confinement of fractional probe
charges, is well-known to occur in
the Schwinger model \cite{js,cks,cjs}.

\section {Higgs phase vs. confinement in the
Schwinger model}
\setcounter{equation}0
First, we consider the well
understood \cite{js,cks,cjs} case of the Schwinger model, with Lagrangian
\begin{equation}
\sL = \bar\psi (i\gamma^\mu \partial_\mu - e \gamma^\mu A_\mu - m)\psi
-{1\over 4} F_{\mu\nu} F^{\mu\nu},
\end{equation}
where
\begin{equation}
F_{\mu\nu}= \partial_\mu A_\nu - \partial_\nu A_\mu =\epsilon_{\mu\nu} F.
\end{equation}
The coupling constant $e$ has dimension of mass.
After bosonizing the Dirac fermion we arrive at the following equivalent
Lagrangian,
\begin{equation}
\sL = {1\over 2} F^2 + {1\over 2} (\partial_\mu \phi)^2
+{e\over \sqrt\pi} F\phi +
m\Sigma [\cos (2\sqrt\pi \phi)-1],
\end{equation}
  where
  \be
 \label{cond}
\Sigma ~=~ e~\frac{\exp(\gamma)}{2\pi^{3/2}}
  \ee
is the absolute value of the fermion condensate in the Schwinger model
($\gamma$ is the Euler constant).
We have added $-1$ to the cosine  so as to have zero classical vacuum energy.

For our purposes it is convenient to integrate out $\phi$ (or
equivalently $\psi$) in order to derive the effective action for
the gauge field. This is particularly easy in the case of  $m=0$, where
the theory is quadratic in $\phi$, and we obtain
\begin{equation}
\sL_{eff} = {1\over 2} F^2 + {e^2\over 2\pi} F {1\over\partial ^2} F.
\end{equation}
The non-local term acts essentially as a mass term for the gauge field.
To show this, we pick $A_1=0$ gauge and restrict ourselves to static
fields, so that $1/\partial^2$ may be replaced by $-1/\partial_1^2$.
After an integration by parts the effective Lagrangian reduces to
\begin{equation}
\sL_{eff} = {1\over 2} (\partial_1 A_0)^2 + {e^2\over 2\pi} A_0^2.
\end{equation}
This may be interpreted as a peculiar two-dimensional version of
the Higgs phenomenon: the Coulomb force is replaced by a force of finite
range with a mass scale $\mu = e/\sqrt \pi$.
The consequences of this may be probed by
introducing a static external charge distribution $\rho(x^1)$.
This adds $-\rho A_0$ to $\sL_{eff}$, and the equation of motion becomes
\begin{equation}
\partial_1^2 A_0 - \mu^2 A_0 = -\rho (x^1).
\label{massive}\end{equation}
Suppose, for instance, that we fix an external charge $e'$ at $x^1=0$,
and $-e'$ at $x^1=a$.
Solving (\ref{massive}) with
\begin{equation}
\label{source}
\rho(x^1) = e' \big (\delta(x^1) - \delta(x^1-a) \big ),
\end{equation}
we get
\begin{equation}
A_0 (x^1) = {e'\over {2\mu}} \big (e^{-\mu |x^1|} - e^{-\mu|x^1-a|}\big ).
\end{equation}
Substituting this back into $\sL_{eff}$ we find that the energy
of the two test charges is
\begin{equation}
\label{Va}
V(a) = {e'^2\over {2\mu}} \big (1- e^{-\mu a}\big ).
\end{equation}
While $V(a)$ increases linearly for small $a$, it saturates at
$e'^2/(2\mu)$ for large separations. This indicates a remarkable
phenomenon: {\sl  any fractional charge $e'$ is screened by integer massless
charges.} Does this also occur when the dynamical charges are massive ?
One way to find the answer is to integrate out $\phi$.
The fact that the massive theory is non-polynomial in
$\phi$ leads to a non-polynomial effective action for $F$.
The expansion of $\sL_{eff} $ in powers of $F$ may be constructed by
integrating
$\phi$ out order by order in $eF$,
\begin{equation}
\label{full}
\sL_{eff} = {1\over 2} F^2 +
{e^2\over 2\pi} F {1\over\partial ^2+
4\pi m\Sigma } F
+ {16 m e^4\over\pi} \left [{1\over\partial ^2+ 4\pi
m\Sigma } F \right ]^4 + \sO(F^6).
\end{equation}
For weak, slowly varying fields this may be approximated by
\begin{equation}
\label{trunc}
\sL_{eff} = {1\over 2} F^2 \left (1+ {e^2\over 4 \pi^2 m\Sigma }\right ).
\end{equation}
Thus, the leading effect of integrating out a massive fermion is
a finite renormalization of electric charge: the Higgs phenomenon has
disappeared. The absence of a mass term for the gauge field means
that we can no longer screen a fractional charge by integer charges.
In other words, $V(a)\sim a$ as $a\rightarrow\infty$, and the theory
is in the confining phase.

Solving the equations of motion which follow from the truncated Lagrangian
(\ref{trunc}) with the source (\ref{source}) and calculating the energy, we
get for small $m \ll e$
  \be
V(a) \ =\ \left( \frac{e'}{e} \right)^2 2\pi^2 m \Sigma a .
 \ee
This is true, however, only as long as $e' \ll e$.
Otherwise, the higher-order
terms in the effective Lagrangian (\ref{full}) cannot be neglected and
the string tension is renormalized. In the following
we determine the exact dependence of the string
tension on $e'/e$ and show that
it vanishes for {\it integer} probe charges. For fractional
probe charges, it vanishes only when $m = 0$, but does {\it not} vanish in
the massive Schwinger model.

One of the ways to reach this conclusion is by studying classical solutions of
the bosonized equations, as in \cite{Ellis}. Let us make the fermions of charge
$e' = qe$ and large mass $M$ dynamical and bosonize them in terms of a new
scalar field $\chi$. The complete Schwinger model Lagrangian becomes
\footnote{We suppress the $\theta$-angle here but find it useful to
introduce it in section 5.}
\begin{equation}
\sL = {1\over 2} F^2 + {1\over 2} (\partial_\mu \phi)^2
+ {1\over 2} (\partial_\mu \chi)^2
+{e\over \sqrt\pi} F(\phi+ q\chi)
+ m\Sigma [\cos (2\sqrt\pi \phi)-1]
+ cM^2 [\cos (2\sqrt\pi \chi)-1] \ ,
\end{equation}
where $c$ is a numerical constant.
After integrating out the gauge field, we arrive at the following
Lagrangian
\begin{equation}
{1\over 2} (\partial_\mu \phi)^2+ {1\over 2} (\partial_\mu \chi)^2
-{e^2\over 2\pi} (\phi+ q\chi)^2
+ m\Sigma [\cos (2\sqrt\pi \phi)-1]
+ cM^2 [\cos (2\sqrt\pi \chi)-1].
\end{equation}
Following \cite{Ellis} we may look for static solutions, $\phi(x^1)$,
$\chi(x^1)$, to the
resulting equations of motion. The requirement of finite energy
leads to the boundary conditions $\phi(-\infty)=\chi(-\infty)=0$.
$\phi$ and $\chi$ must also approach constant values as $x^1\rightarrow
\infty$. For $m=0$ there exists a finite energy solution
with
\begin{equation}
\chi(\infty)=\sqrt\pi\ ,\qquad\qquad \phi(\infty) =-q\sqrt\pi
\ .
\end{equation}
The total charge,
\begin{equation}
Q={e\over\sqrt\pi} [\phi(\infty)-\phi(-\infty)]
+{e'\over\sqrt\pi} [\chi(\infty)-\chi(-\infty)]\ ,
\end{equation}
vanishes for such a solution, as it should. This solution describes
a massive charge $e'$ screened by a cloud of massless charges $e$.
It provides us with a rather detailed understanding of the mechanism
for this screening. The bosonized theory with a massless field
$\phi$ possesses finite energy configurations containing any desired
charge $-e'$ in a localized region of space. Upon gauging
of the theory, these configurations
bind to charge $e'$ and neutralize it. Remarkably, such fractionally
charged $\phi$-solitons acquire infinite energy as soon as $m$ is turned
on, due to the $m\Sigma [\cos (2\sqrt\pi \phi)-1]$
term in $\sL$. For small $m$, the energy per unit length (i.e. the string
tension)
may be found from the first order perturbation theory and is given by
\begin{equation}
\label{tension}
\sigma = m\Sigma [1-\cos (2\pi q )].
\end{equation}
In section 4 this result will be rederived by analyzing the
 behavior  of the  Wilson loop in the path integral approach.

We see that the string tension indeed vanishes when $e'$ is an integer
multiple of $e$. This has an obvious physical interpretation:
one can always screen an integer charge by binding to it a number of particles
of charge $-e$.

\section{Non-abelian Higgs phase in the
massless adjoint fermion model}

\setcounter{equation}0
In the previous section we discussed the Schwinger
model. As is well-known \cite{js,cks,cjs},
it is in the Higgs phase for massless fermions and,
for fractional probe charges, in
the confining phase for massive fermions. In this section we show that
essentially the same conclusions hold in certain non-Abelian
1+1 dimensional gauge theories.

While in 3+1 dimensions confinement is a rather miraculous phenomenon,
which is not yet fully understood, in 1+1 dimensions it is hardly a
mystery due to the confining nature of the Coulomb force \cite{ea}.
In a pure $SU(N)$ gauge theory, for example, there are no
dynamical gluons, but there exists an exactly linear Coulomb
potential between test charges. In other words, Wilson loops in
any representation will exhibit an area law.

If, however, we couple dynamical fermions in the fundamental
representation to the gauge field, then the situation changes.
The Wilson loops in the fundamental (or any other) representation
now exhibit the perimeter law because the dynamical fundamental charges
screen the test charges.

A more interesting situation is expected
to occur in theories where all the dynamical fields are in the adjoint
representation of $SU(N)$. Such 1+1 dimensional models have received
some recent attention because of their many similarities with
3+1 dimensional gauge theories \cite{DK,Kutasov,num}.
The adjoint fields play a physical role similar to that of
transverse gluons. In theories where all the dynamical fields are in
the adjoint representation the adjoint Wilson loop exhibits the perimeter law,
while the Wilson loop in the fundamental representation is usually
expected to obey the area law corresponding to confinement.
 It is interesting that in 1+1 dimensional models with
adjoint matter the criteria for confinement are the same as in
the 3+1 dimensional gauge theories.
The proof of confinement is expected to be much simpler in 1+1 dimensions.
To our surprise, however, we will find non-Abelian models where the
confining phase is replaced by the Higgs phase, much like in the Schwinger
model. In this section we discuss the simplest such model: SU(N) gauge
theory coupled to a massless Majorana fermion in the adjoint representation.
We will show that test charges in the fundamental representation are
screened by the massless adjoint fermions. This is the non-Abelian analogue
of the screening of fractional charge by massless integer charges that we
observed in the Schwinger model.

The gauged Lagrangian for a single flavor of massless Majorana fermions is
\begin{equation}
\sL = \tr\left [i\bar\psi \gamma^\mu D_\mu \psi
-{1\over 4 g^2} F_{\mu\nu} F^{\mu\nu} \right ]\ ,
\end{equation}
where $\psi=\psi^a t^a$, $A_\mu= A_\mu^a t^a$, and $t^a$ are the
$N^2-1$ hermitian generators of $SU(N)$. The field strength and covariant
derivative are defined as
\begin{eqnarray}
& F_{\mu\nu}= \partial_\mu A_\nu - \partial_\nu A_\mu
+ i[A_\mu, A_\nu]=\epsilon_{\mu\nu} F \ ,& \\
&D_\mu \psi =\partial_\mu \psi + i[A_\mu, \psi]\ . &
\end{eqnarray}
We will integrate out $\psi$ to derive the effective action for
$A_\mu$ which is known explicitly in 1+1 dimensions \cite{PW,P},
\begin{equation}
S_{eff} =\tr \int d^2 x \left [{1\over 2 g^2} F^2 +
{N\over 2\pi} (\partial_- A_+ - \partial_+ A_-) {1\over\partial ^2}
(\partial_- A_+ - \partial_+ A_-) \right ] + N S_{WZ}(A_+) - N S_{WZ}(A_-).
\label{nonloc}\end{equation}
The non-local Wess-Zumino term is an integral over manifold $B$ whose boundary
is space-time,
\begin{eqnarray}
S_{WZ}(A_+)&={1\over 12\pi}\int_B d^3 x \epsilon^{ijk}
\partial_i g g^{-1} \partial_j g g^{-1} \partial_k g g^{-1}\ , \\
S_{WZ}(A_-)&={1\over 12\pi}\int_B d^3 x \epsilon^{ijk}
\partial_i h h^{-1} \partial_j h h^{-1} \partial_k h h^{-1}\ ,
\end{eqnarray}
where $A_+ = \partial_+ g g^{-1}$ and $A_- = \partial_- h h^{-1}$.
The factor of $N$ in the induced action is the central charge of
the affine algebra of the $SU(N)$ gauge currents.
In \cite{ks} it was noted that
an identical current algebra results in the gauged model of $N$
flavors of massless Dirac fermions in the fundamental representation
of $SU(N)$ (theory II). It was further shown \cite{ks} that
the massive spectrum of theory II is identical to that of
theory I (gauge theory coupled to
one massless adjoint multiplet). Theories
I and II are not completely equivalent because I has no massless bound states
while II does, but the massless sector in II is
in some sense decoupled from the rest of the
spectrum.
The fact most important for us is that, since their gauge current
algebras are identical, theories I and II
have identical effective actions for $A_\mu$.
\footnote{Cf. the abelian case: the
Schwinger model with four massless fermions of charge
$e/2$ has the same effective action as the theory with one massless fermion of
charge $e$.}
As a result, the expectation value of any Wilson loop,
\begin{equation}
\langle W\rangle = \int [\sD A]\ W {\rm e}^{- S_{E}(A)} ,
\end{equation}
is the same in theories I and II ($S_{E}(A)$ is the Euclidean
continuation of $S_{eff}(A)$).
It is physically clear that in the model with massless fundamental fermions
(II) the Wilson loops in the fundamental (and all other) representations
must obey the perimeter law: the fundamental fermions can screen
test charge in any representation. This implies that in theory I the
fundamental Wilson loop also obeys the perimeter law. Surprisingly,
we have shown that the theory with a massless adjoint Majorana multiplet is
not confining: it is rather in the Higgs phase.
In the following we will confirm this unexpected conclusion in a number
of ways.

There is a
subtlety in the above argument that requires further explanation.
Theory I has gauge group $SU(N)/Z_N$. As we explain in detail in section 4,
there are $N$ different
topological classes for $A_\mu$ associated with
the elements of $\Pi_1 (SU(N)/Z_N)=Z_N$.
Only one of them, the trivial class,
is present in theory II.
However, as explained in section 4, each of the topologically non-trivial
classes in I has fermion zero modes and does not contribute to
$\langle W\rangle$. We expect, therefore, that $S_E(A)$, the Euclidean
effective action obtained by integrating the fermions out, diverges
for the topologically non-trivial configurations.
To show this, let us consider the Euclidean theory defined on $S^2$.
In the topologically non-trivial sectors $A_\mu$ is not single-valued.
It has singularities of the Dirac string type where the infinitesimal
Wilson loop surrounding the north pole is a non-trivial element of
$Z_N$. It is not hard to show that such a singularity creates a
divergence in the effective action (\ref{nonloc}).
For simplicity, we consider $N=2$, but the argument generalizes to
other $N$. Near the north pole (as $r\rightarrow 0$)
\begin{equation}
A_\mu\rightarrow i \Omega^\dagger \partial_\mu \Omega.
\end{equation}
In the instanton configuration we may choose a gauge where
\begin{equation}
\Omega\rightarrow \exp (i \theta\sigma_3/2),
\end{equation}
so that only $A_\mu^3$ is not single valued and
\begin{equation}
F^3 = \pi \delta^2 (x) +{\rm regular\ terms}.
\end{equation}
Thus, the effective action in the instanton class diverges
due to the term
\begin{equation}
- 2\pi \int d^2 x \delta^2 (x) {1\over \partial^2} \delta^2 (x).
\end{equation}
This is the well-known expression for the electrostatic self-energy
of a two-dimensional charge, which is logarithmically divergent.
The fact that $S_E(A)$ turns out to be infinitely
large in the instanton sectors is directly related to the presence
of the fermion zero modes before the fermions are integrated out.
\footnote {Strictly speaking, this reasoning is not quite rigourous. To be
precise, one should treat the topologically distinct sectors separately and
single out the contribution of zero modes explicitly (see Ref.\cite{Wipf} for a
detailed analysis in the Schwinger model case). The
more precise treatment of
topologically non-trivial sectors in $QCD_2$ with adjoint fermions will be
given in Sect.5, but it is rather remarkable
 that  heuristic arguments based on the universal form of the
effective action give essentially the same answer.}
The divergence of $S_E(A)$
suppresses the instantons in theory I and restores
its equivalence to theory II.
As a result, all Wilson loops in I and II are identical.

One interesting check of the screening phenomenon
involves a calculation of the static quark--antiquark
potential. The charge of the quark and the antiquark points in one of the
$N^2-1$ directions of $SU(N)$, which we call direction 1 without any
loss of generality,
\begin{equation}
\rho^1 (x)\sim \delta(x)- \delta(x-a)\ ,\qquad\qquad
\rho^a (x)=0\ , \ a\neq 1.
\end{equation}
We will choose the $A_1=0$ gauge and
look for a static
classical solution for $A_0$ in the background of this charge
density. The classical gauge field points in the same group direction as
the charge density, $A_0^a=0$ for $a\neq 1$. The Wess-Zumino terms may be
neglected because they involve group commutators, while $g$, $h$ and their
derivatives commute. Thus, the equations satisfies by a static
$A_0^1$ are the same as in the Abelian theory,
\begin{equation}
\partial_x^2 A_0^1 - {g^2 N\over \pi} A_0^1 = g\sqrt N (\delta(x-a)
-\delta(x))
\ .\end{equation}
Note that the screening mass-squared is $\mu^2= g^2 N/\pi$, which is finite in
the large $N$ limit. Substituting the solution into the effective action,
we find that the static quark--antiquark potential behaves as
\begin{equation}
V(a)\sim \mu (1- e^{-\mu a}).
\label{saddle}\end{equation}
In the Schwinger model, where the effective action for $A_\mu$ was quadratic,
this method of calculation was exact. In the non-Abelian case we
may only hope to have found the dominant saddle point. The fluctuations around
it probably change the simple formula (\ref{saddle}) but do not alter
its qualitative behavior, which is  characteristic of the Higgs phase.

At this point it is interesting to ask how the transition from
confinement to screening affects the spectrum of the theory.
In the large $N$ limit,
the spectrum of single-``glueball'' states is expected to be fully
discrete
for any non-vanishing fermion mass. For highly excited states,
however, the gaps become astronomically small due to the exponentially
growing density of states. This is the kind of structure one expects
to find in physically interesting confining gauge theories.
As we have argued above, for $m=0$ confinement is replaced by screening.
The disappearance of the string tension may lead to a continuous
spectrum, at least for high enough excitation number.
\footnote{While in a theory with the adjoint matter alone the continuous
spectrum is only a hypothesis, we can do better for a theory with both
a massless adjoint and a fundamental fermion mutliplets.
In addition to the glueball-like states this theory contains mesons
(open strings) with the fundamental fermions at the end-points.
The large $N$ spectrum
of such mesons should become continuous at the energy sufficient for
a decay into a quark screened by a
cloud of massless adjoint quanta and an antiquark
screened by a cloud of massless adjoint quanta.
Thus we expect a meson spectrum consisting perhaps
of a few low-lying discrete
states followed by a continuum.}
In the numerical work of \cite{num} the lowest couple of states
were found to have discrete gaps, but beyond that it was difficult to
judge whether the spectrum is continuous or discrete with very small
gaps.
It is necessary to improve numerical techniques to the point where
it is possible to judge whether
the transition to continuous spectrum takes place.
If it does, then it is clearly interesting to identify the precise energy
where the spectrum becomes continuous.

\section{Changing the group representation via bosonization.}
\setcounter{equation}0
To make the screening of fundamental charges by massless adjoint fermions
less mysterious we should identify the operators in the free
fermion theory which transform in the fundamental representation
of $SU(N)$. Such operators are analogous to the fractionally charged
solitons of the massless bosonized field which, as we showed in the
previous section, screen elementary fractional charges in the
Schwinger model. Below we sketch a similar construction in the
simplest adjoint fermion model corresponding to $SU(2)$.

We will consider the left-moving (holomorphic) sector of the free fermion
theory (the antiholomorphic sector behaves analogously).
The fermion fields $\psi^a (z)$, $a=1, 2, 3$, transform in the adjoint
(triplet) representation under the $SU(2)$ currents
\begin{equation}
J^a (z) = {i\over 2} \epsilon^{abc} \psi^b \psi^c.
\end{equation}
It is convenient to combine $\psi^1$ and $\psi^2$ into a Dirac fermion,
which may be bosonized in terms of the holomorphic part of a boson field,
\begin{equation}
\psi^1+i\psi^2= \sqrt 2
e^{i\phi (z)}\ , \qquad\qquad \psi^1-i\psi^2=\sqrt 2  e^{-i\phi (z)}.
\end{equation}
The currents assume the form
\begin{eqnarray}
& J^+= J^1+i J^2= \sqrt 2 \psi^3 e^{i\phi}\ ,& \qquad
J^3 = -i\partial_z \phi \ , \\
& J^-= J^1-i J^2= \sqrt 2 \psi^3 e^{-i\phi}\ .&
\end{eqnarray}
Let us recall that the $c=1/2$ theory corresponding to $\psi^3$ contains
order and disorder operators with the following OPE,
\begin{equation}
\psi^3 (z) \sigma (0) =-{1\over \sqrt {2z}} \mu(0)\ , \qquad
\psi^3 (z) \mu (0) =-{1\over \sqrt {2z}} \sigma (0)\ .
\end{equation}
Now it is not hard to see that the operators
\begin{equation}
\Psi_+ = \sigma e^{i\phi/2}\ ,\qquad \Psi_- = \mu e^{-i\phi/2}
\end{equation}
are local with respect to the $SU(2)$ currents and, in fact,
transform in the fundamental (doublet) representation,
\begin{eqnarray}
&J^3 (z) \Psi_+ (0) =-{1\over 2 z} \Psi_+ (0)\ ,\qquad
J^3 (z) \Psi_- (0) ={1\over 2 z} \Psi_- (0)\ , &\\
&J^+ (z) \Psi_- (0) =-{1\over z} \Psi_+ (0)\ ,
\qquad J^- (z) \Psi_+ (0) =-{1\over z} \Psi_- (0)\ ,
&\end{eqnarray}
where we have exhibited only the singular terms in the OPE.
The doublet fields have a rather exotic holomorphic dimension, $3/16$
(this is the sum of the holomorphic dimension of $\mu$ or $\sigma$,
$1/16$, and that of $e^{\pm i\phi/2}$, which is $1/8$).
This is not too surprising because in the bosonized theory of a single
massless Dirac fermion
the fractionally charged objects, $e^{i q\phi}$,
also have fractional dimensions, $q^2/2$.
Nevertheless, it is these objects that screen external
fractional static charges  in the Schwinger model.

It is thus plausible that the composite doublets
we found in the free adjoint theory are capable of screening the external
test doublets in the gauged theory. We believe, although have not checked
in detail, that similar constructions of fundamentals from adjoints are
possible for all $SU(N)$ gauge groups. In fact, somewhat simpler
constructions of a similar type demonstrate the screening of
external spinor charges in $SO(2n)$ gauge theory with massless fermions
in the vector representation.

Consider, for instance, the $SO(8)$ gauge theory
with fermion fields $\psi^a (z)$, $a=1, 2, \ldots, 8$,
transforming as a vector. We may combine the 8 Majorana
fermions into 4 Dirac fermions and bosonize them
\begin{equation}
\psi^1+i\psi^2= \sqrt 2 C_1
e^{i\phi_1 (z)}\ , \qquad\qquad \psi^3+i\psi^4=\sqrt 2 C_2 e^{i\phi_2 (z)}
\ ,
\nonumber\end{equation}
\begin{equation}
\psi^5+i\psi^6= \sqrt 2 C_3
e^{i\phi_3 (z)}\ , \qquad\qquad \psi^7+i\psi^8=\sqrt 2 C_4 e^{i\phi_4 (z)}
\ ,
\end{equation}
where $C_i$ are the cocycle operators necessary for maintaining the
proper anticommutation relations between different fermion fields.
As is well known in string theory \cite{gsw},
the fields that transform as spinors of $SO(8)$ may be easily
constructed out of the bosonic fields as
\begin{equation}
\tilde C\exp i\left (\pm {\phi_1\over 2} \pm {\phi_2\over 2}\pm {\phi_3\over 2}
\pm {\phi_4\over 2}\right )
\end{equation}
where $\tilde C$ are the necessary cocycles. The chirality of the spinor is
the product of the signs that appear in the exponent.
The special property of the $SO(8)$ is that
the objects that transform as spinors have dimension $1/2$ which
means that they are fermions, just like the original fields that
transform as a vector. This is not surprising because
the two spinors (of positive and negative chirality) and the
vector are interchanged by the triality of $SO(8)$.
We conclude that there exists an exact transformation that
maps the gauge theory with massless fermions that transform as a vector
of $SO(8)$ and into the theory of massless fermions that transform as
a spinor of definite chirality (we are free to chose whether it is
positive or negative). This tranformation preserves the number of
fermion fields. To show that the external static spinor charges are
screened rather than confined we simply perform the transformation
on the lagrangian.
Thus, it is certain that the ``composite'' spinor fermions
screen the external static spinor charges.
While for other groups the ``composite'' objects have more
exotic dimensions, it is still very plausible that they screen external
static charges that transform in different representation from those
appearing in the lagrangian.

Our ability to carry out constructions such as those shown above
depends crucially on special
properties of conformal field theories. Once the mass is turned on,
the fractionally charged solitons of the type we used no longer exist.
Thus, we expect that external fundamental charges can no longer be screened
and we have confinement.
Examination of the quadratic terms in $S_{eff}(A)$ for $m\neq 0$
also indicates that the mass term for the gauge field is no longer present.
For small $m$ we expect the theory to be confining, with a small string
tension. As we show in the next section, this is indeed what happens.

\section{Wilson loops and the topological structure.}
\setcounter{equation}0
\subsection{The Schwinger model}
 Consider first
the Wilson loop with unit probe charge. In the
massless Schwinger model the functional integral is
Gaussian, and the higher-order correlators factorize into
products of pair correlators. Therefore, we find
  \be
  \label{WFF}
\left< {\rm e}^{ie \int_C A_\mu dx_\mu} \right> =
\left< {\rm e}^{ie \int_D F(x) d^2x} \right> = \nonumber \\
\exp \left\{ - \frac 12 e^2 \int_D \int_D d^2x d^2y <F(x) F(y)> \right\}.
  \ee
 The correlator $<F(x) F(y)>$ has the form (see e.g.
\cite{Wipf})
  \be
  \label{FF}
<F(x) F(y)> = \delta(x-y) - \frac {\mu^2}{2\pi} K_0(\mu|x-y|),
  \ee
where $\mu^2=e^2/\pi$.
It satisfies the property
  \be
  \label{susc0}
 \int d^2x <F(x) F(0)> \ = \ 0.
\ee
The property (\ref{susc0}) is natural, of course. In the Schwinger model, $F(x)
$ is the local density of the topological charge:
  \be
  \label{nudef}
\nu = \frac e{2\pi} \int F(x) d^2x.
  \ee
The integral on the LHS of Eq. (\ref{susc0}) is proportional to
the topological susceptibility
  \be
  \label{chi}
\chi = \frac 1V <\nu^2> = \left( \frac e{2\pi}\right)^2 \int d^2x <F(x) F(0)>,
  \ee
which is zero in the theory with massless fermions: the topologically
non-trivial sectors with $\nu \neq 0$ involve fermion zero modes which
make the corresponding contributions to the partition
function vanish.

Note that in the quenched Schwinger model (without dynamical fermions) the
correlator $<F(x) F(y)>$ is just $\delta(x-y)$  and the topological
susceptibility (\ref{chi}) is not zero
(cf. the well-known situation in 4D Yang-Mills theory: the topological
susceptibility is zero in $QCD_4$ with massless quarks, but has a
non-zero value $\chi_{YM} \sim \Lambda_{YM}^4$ in the pure Yang-Mills theory).

The property (\ref{susc0}) leads to the vanishing of the coefficient of the
area in $\ln <W(C)>$, and the Wilson loop has the perimeter law
  \be
  \label{perim}
<W(C)> \sim \exp \left\{ - e^2P/(4\mu) \right \}
\ee
for large contours.\footnote{The coefficient of $P$
may be calculated by doing the integral in
(\ref{WFF}) with the
account of boundary effects  \cite{vacSM} or, alternatively,
from (\ref{Va}) after taking
the limit $a \rightarrow \infty. $}
 Static heavy charged sources are screened by the massless
dynamical fermions. In the quenched Schwinger model, the susceptibility
(\ref{chi}) is non-zero, and the Wilson loop has the area law corresponding
to the linearly rising static Coulomb potential.

Let us consider now the Wilson loop for a fractional probe charge $e' = qe$
\be
 <W_q(C)> = \left< \exp \left\{ ieq \int_C A_\mu dx_\mu \right
\} \right>
 \ee
The derivation presented above can be
easily generalized to this
case, and we find that $<W_q(C)>$ displays
the perimeter law, i.e. the dynamical
fermions with integer charges somehow manage to
screen a heavy  probe of arbitrary charge.
This fact has been noted by many people and is a common
lore. The mechanism of this strange screening deserves some further
explanation, however.

Let us note that the perimeter law holds for the
integer $q$ Wilson loops even after the
fermions are endowed with a mass. However, for non-integer
$q$, $<W_q(C)>$
exhibits the area law behavior corresponding to confinement for any
non-zero $m$, however small it is \cite{cjs}.
This was already shown in section 3 using bosonization, but here we give
an independent derivation of this remarkable phenomenon.

Note first of all that the topological susceptibility
(\ref{chi}) is no longer zero when $m \neq 0$. For $m \ll e$
it can be calculated exactly. The quickest way to find it is
by introducing
the vacuum angle $\theta$ (this adds the term $i\nu \theta$
to the Euclidean Lagrangian). Then the relation
  \be
  \label{chithet}
\chi =  \frac{\partial^2 \epsilon_{vac}(\theta)}{\partial \theta^2}
\left|_{\theta = 0} \right.
  \ee
holds ($\epsilon_{vac}$ is the vacuum energy density). Consider the
function $\epsilon_{vac} (m, \theta)$ where $m$ can be complex in general.
The point is that $\epsilon_{vac}$ is not an arbitrary function, but rather a
function of a
single complex variable $z = m e^{i\theta}$
(and its complex conjugate) \cite{cjs}.
This follows from the Ward identities and the topological structure of the
theory, and can be derived in the same way as in $QCD_4$.
When $m$ is small and real, we can expand the real
function $\epsilon_{vac}(z, \bar z)$ in a Taylor series,
  \be
  \label{Evac}
\epsilon_{vac} = \epsilon_{vac}(0) - \frac 12 \Sigma (z + \bar z) + {\cal
O}(z^2)
= \epsilon_{vac}(0) - \Sigma m \cos \theta + {\cal O} (m^2).
  \ee
The quantity $-\Sigma$ is simply the fermion condensate
at  $\theta = 0$ as given in Eq.(\ref{cond}):
  \be
  \label{condSM}
<\bar \psi^a \psi^a>_{\theta = 0}\  =\ \frac \partial {\partial
m}\epsilon_{vac}(m)
\left|_{m=\theta=0} = -\Sigma .  \right.
 \ee
Substituting (\ref{Evac}) into (\ref{chithet}), we immediately get the
relation
  \be
  \label{chisigm}
\chi = \Sigma m .
  \ee
Again, this relation is analogous to the well-known relation
$\chi = \Sigma m /N_f$ in $QCD_4$ derived in \cite{Crewt} (see also
\cite{Leut} ).

If (\ref{chisigm}) is substituted into (\ref{WFF}), then the string
tension is found to be non-vanishing (and proportional to $m$) for any
probe charge $q$.
This is wrong, however. The point is that the {\it massive} Schwinger model
is no longer  a Gaussian,  exactly soluble,  theory.
The higher-order correlators
no longer factorize into products of pair correlators but involve
non-trivial connected pieces. For the $q=1$ Wilson loop one can write
  \be
  \label{W1C}
<W_1(C)> = \exp \left\{ - \frac 12 e^2 \int_D \int_D d^2x d^2y <F(x) F(y)> +
\right. \nonumber \\
\left. \frac{e^4}{24} \int_D \int_D \int_D \int_D d^2x d^2y d^2z d^2u
<F(x) F(y) F(z) F(u)>_c \ -\ \ldots \right\}.
  \ee
 Since we are interested only in the
coefficient of the area in $\ln~ <W_1(C)>$ for large contours, Eq.(\ref{W1C})
can be rewritten as
  \be
  \label{W1Cchi}
<W_1(C)> = \exp \left \{ {\cal A}_D \sum_{n=1}^\infty (-1)^n \frac {(2\pi)
^{2n}}{
(2n)!} \chi_{2n} \right \},
  \ee
where
  \be
\label{chi2n}
\chi_{2n} = (-1)^{n+1}  \frac{\partial^{2n} \epsilon_{vac}(\theta)}
{(\partial \theta)^{2n}} =
(\frac e{2\pi})^{2n} \prod_{i=1}^{2n-1}\int d^2 x_i
<F(0) F(x_1) \cdots F(x_{2n-1})>_c,
  \ee
are the generalized susceptibilities. For $m \ll e$ they can be easily
found from (\ref{Evac}), and we get for the string tension
  \be
  \label{sig1}
\sigma = - {1\over{\cal A}_D} \ln <W_1(C)> = \Sigma m (1 - \cos 2\pi ) = 0.
  \ee
This can be easily understood by noting that, if one is interested only
in the coefficient of the area, the integral
in $<\exp\{ie \int d^2x F(x)\}>$ can be extended over the whole two-
dimensional manifold where the theory is defined (the manifold may be very
large but compact to provide for infrared regularization of the path
integral). This is implicit in (\ref{chi2n}).
The flux of the electric field through the area has the meaning of the net
topological charge (\ref{nudef}) on the whole manifold. We thus have  that
  \be
  \label{W1cnu}
<W_1^{asympt} (C)> = \left< e^{2\pi i \nu} \right>,
 \ee
and, since $\nu$ is quantized to be an integer, this is
manifestly equal to 1 (the perimeter
corrections are due to the boundary effects in the flux integral and are
disregarded in this reasoning).

Now consider the Wilson loop of arbitrary test charge $q$,
$W_q(C)$. Eqs. (\ref{W1C}-\ref{sig1}) are easily generalized, and one
finds that the string tension is
  \be
  \label{sigq}
\sigma = - {1\over{\cal A}_D} \ln <W_q(C)> = \Sigma m (1 - \cos (2\pi q)).
  \ee
This result was obtained earlier via bosonization,
and it can also be understood as follows.
In the limit where the boundary effects are neglected,
we get
  \be
  \label{Wqnu}
<W_q^{asympt} (C)> = \left< e^{2\pi iq \nu} \right>.
  \ee
The value of $e^{2\pi iq\nu}$ depends on $\nu$ and the average
is not 1 anymore. Actually, (\ref{Wqnu}) can be written as
  \be
  \label{Zratio}
<W_q^{asympt} (C)> = \  \frac{Z(\theta = 2\pi q)}{Z(\theta = 0)},
  \ee
where
  \be
  \label{Zthet}
Z(\theta) \equiv \sum_\nu Z_\nu e^{i\nu\theta} \ =\ \exp\{-\epsilon_{vac}
(\theta) {\cal A} \},
  \ee
and ${\cal A}$ is the total area. Substituting (\ref{Evac}), we
immediately find (\ref{sigq}).
The string tension goes to zero and confinement disappears in the limit
$m \rightarrow 0$. Again, this can be easily understood from the
representation (\ref{Zratio})
and the Fourier decomposition for $Z(\theta)$.
For massless fermions, only the trivial topological sector with $\nu = 0$
contributes to the partition function. The contribution of the non-trivial
sectors is killed by the fermion zero modes which appear due to the index
theorem. Thus, $Z$ is $\theta$-independent and $<W_q^{asympt}(C)>|_{m=0}\  =
\ 1$.

\subsection{$QCD_2$ with adjoint fermions.}

The behavior of the
Wilson loop may be related to the topological structure of the
theory also in $QCD_2$ with adjoint fermions. It was observed in
\cite{Witten} and shown in detail in \cite{adjZN}, and later using the
Hamiltonian formalism in \cite{Lenz}, that the adjoint $QCD_2$ has $N$
distinct topological classes for Euclidean gauge field configurations.
This is because the true gauge group in this theory is $SU(N)/
Z_N$ rather than $SU(N)$ (the adjoint fields are not transformed under
the action of the center), and $\pi_1[SU(N)/Z_N] = Z_N$ is
non-trivial.
If we define the theory on a Euclidean plane, for instance, then
the admissible boundary conditions are
\begin{equation}
\lim_{r\rightarrow\infty} A_\mu=i \Omega^\dagger \partial_\mu \Omega,
\end{equation}
where $\Omega\in SU(N)/Z_N$. There are $N$ topologically distinct ways
to map the circle at infinity into $SU(N)/Z_N$. Therefore, there are
$N$ distinct topological classes for $A_\mu$.

Consider first
the well-understood case $N = 2$. The gauge group is $SU(2)/Z_2
= SO(3)$. There are just two topological classes --- the trivial class  and the
class containing one instanton. One can be convinced \cite{Witten} that for all
the topologically trivial fields
  \be
 \label{Wnab}
W(C)=
\frac 12 {\rm Tr \ P} \exp \left\{ ig \int_C A_\mu^a t^a dx_\mu \right\} \ =
\ 1 ,
  \ee
and for the non-trivial fields $W(C)=-1$. The contour $C$ runs around infinity
on the Euclidean plane.
Alternatively we may compactify the Euclidean space to $S^2$ by,
say, the stereographic projection. Then the contour $C$ surrounds
the north pole of the sphere where the field $A_\mu(x)$
is pure gauge $i\Omega^\dagger(x) \partial_\mu \Omega(x)$ with a trivial or
non-trivial mapping $S^1 \rightarrow SO(3)$.

The average of (\ref{Wnab}) is the order parameter for
the screening or the
confinement phase. The loop $C$ in that case should
be large but not necessarily surrounding the whole two-dimensional Euclidean
manifold. However, as we have seen when discussing
the Schwinger model, since
we are interested only in the string tension, it is
sufficient to study $<W(C)>$ for loops at infinity.

In the hamiltonian language, there are 2 classical vacua related by a
topologically non-trivial {\it large} gauge transformation, and a
superselection rule which is quite analogous to the standard $\theta$-angle
superselection rule in $QCD$ \cite{theta} may be imposed. The only difference
is that here there are only two possible values of
$\theta$: $\theta = 0$ and $\theta
= \pi$. The partition function in these two sectors has the form
  \be
  \label{Zpm}
Z_\pm = Z_{triv} \pm Z_{inst}.
  \ee
The crucial observation is that any gauge field in the instanton sector
involves 2 fermion zero modes \cite{adjZN}, which implies
that the expectation value of $W(C)$ is equal to its value in the
topologically trivial
sector, $\langle W(C)\rangle=1$. Therefore, the
string tension is zero and we are in the screening or Higgs phase, in
accordance with what we argued in section 3.

The appearance of the fermion zero modes
in the non-abelian 2D instanton background
is not as straightforward
as in the abelian case because we do not have an index
theorem of the Atiyah-Singer kind: the topological charge cannot be presented
here as an integral of a local charge density. However, the presence of
the zero modes can be seen in
a number of ways. In \cite{adjZN}, they have been
constructed explicitly in a particular instanton gauge field configuration
$A_\mu^{(0)}$ on a torus, and it was shown that they are also there
in a perturbed background $A_\mu^{(0)} + a_\mu$ to all orders in $a_\mu$. In
\cite{Lenz}, the theory was studied in the Hamiltonian approach and the level
crossing phenomenon showed the existence of the zero modes.

Consider $QCD_2$ with adjoint fermions defined on a finite (not necessarily
small) spatial circle of length $L$. Impose the gauge $A^a_0 = 0$. It is
possible to show that the trivial perturbative vacuum $A^a_1 = 0$ has a
gauge copy
  \be
  \label{copy}
 A^a_1 = \ \frac{2\pi}{gL} n^a,\ \ \ \ \ \ \ \ \ \ (n^a)^2 = 1.
  \ee
The field (\ref{copy}) is related to $A^a_1 = 0$ by a {\it large} gauge
transformation not reducible to zero by infinitesimal deformations (the
configurations (\ref{copy}) with different $n^a$ are related to each other
by topologically trivial gauge transformations). Therefore, the {\it
energy} spectrum of the Dirac operator in the background (\ref{copy}) is
exactly the same as for the free operator. Studying the spectrum in the
constant $A^a_1$ background
smoothly interpolating between (\ref{copy}) and
the trivial vacuum,
one can be convinced that one left-handed mode and one right-handed mode
cross zero and the spectrum is rearranged. Therefore,
the level crossing should
occur on {\it any} interpolating path which implies the presence of 2 zero
modes of the Euclidean Dirac operator in any instanton background interpolating
between the inequivalent vacua.

Let us now give the fermion a small mass $m \ll g$. The zero modes in the
instanton sector generate a bilinear fermion condensate
\cite{adjZN}:
  \be
 \label{condNA}
-<\bar \psi^a \psi^a>\ \equiv \ \Sigma \ \sim \ g.
  \ee
There is a gap in the physical spectrum, hence the partition functions
$Z_\pm$ enjoy the extensive property $Z_\pm = \exp\{-\epsilon_\pm(m) {\cal
A}\}$. The fermion condensate is just the first Taylor coefficient in the
expansion of $\epsilon_\pm$ in $m$, and we have for small masses
 \be
 \label{ZpmA}
Z_\pm \sim \exp\{\pm \Sigma m {\cal A}\}.
  \ee
Let us calculate $\langle W(C)\rangle$ in the sector $|+>$. We have
  \be
\langle W_+(C)\rangle =
\frac {Z_{triv} - Z_{inst}}{Z_{triv} + Z_{inst}} = \frac {Z_-} {Z_+}
\sim e^{-2\Sigma m \sA}.
 \ee
Hence in the theory with non-zero Majorana fermion mass, confinement is
restored and the string tension is
 \be
 \label{tensNA}
\sigma = 2\Sigma m.
  \ee

In calculating the string tension for theories
with $N \geq 3$ we encounter a peculiar
difficulty. These theories are in a sense paradoxical and the paradox is
still unresolved. There are $N$ distinct topological sectors, and one
finds that each non-trivial sector involves $2(N-1)$ fermion zero modes.
This number is too large for a bilinear
fermion condensate to be generated. On the
other hand, bosonization arguments suggest that the fermion condensate {\it is}
generated\footnote{Independent arguments show that it is generated in the
infinite
$N_c$ limit \cite{Zhit}.}.
The paradox is akin to a similar controversy which arises in
supersymmetric Yang-Mills theories with higher orthogonal or exceptional
gauge groups \cite{Shif}. These issues are discussed in detail in \cite{adjZN}.

For $m=0$,
however, $\langle W(C)\rangle$
is not sensitive to the exact number of zero
modes in the
topologically non-trivial sectors. The important fact is that the zero
modes {\it are} present and suppress the contribution of all topologically
non-trivial sectors. Thus, $\langle W(C)\rangle$ is simply equal to
its value in the trivial sector, $\langle W(C) \rangle = 1$, and we find a
vanishing string tension. For non-zero mass,
the string tension is not zero anymore, but its dependence on $m$
has not been sorted out.

\subsection{Loop equations in QCD$_2$ with adjoint fermions.}

The screening of the fundamental Wilson loop by massless adjoint fermions
follows also from the loop equations. The idea is to regard the expectation
value of a Wilson loop $W[C]$ as a functional of the contour $C$. Observing
how $W[C]$ changes as we make an infinitesimal variation of $C$ one
obtains a functional differential equation which constrains
the Wilson loop \cite{Migdal}.

To derive this equation, consider the path integral defining the
Wilson loop,
\be
\label{defloop}
W[C]={1\over N}
\int[{\cal D}A_{\mu}][{\cal D}\psi]\thinspace{\rm e}^{-S[A_{\mu}, \psi]}
\thinspace
{\rm tr}\left[{\rm P}\exp\left(i\oint_C A_{\mu}(x)\, dx^{\mu}\right)\right],
\ee
with the action
\be
\label{action}
S[A_{\mu}, \psi]={\rm tr}\int d^2x \left[ i \bar{\psi}\gamma^{\mu}D_{\mu}
\psi -{1\over 4g^2} F_{\mu \nu}F^{\mu \nu}\right].
\ee
Make an infinitesimal change of the integration variables $A_{\mu}(x)\to
A_{\mu}(x)+\delta A_{\mu}(x)$ and $\psi(x)\to \psi(x)+\delta \psi(x)$.
Obviously, this does not change the total path integral. On the other hand,
taken separately, the action and the path ordered exponential do change.
Requiring that these changes balance each other we get a set of
Schwinger--Dyson equations on the Wilson loops---the loop equations.

Together with the path ordered exponentials of the gauge field, such equations
would involve correlators of fermions. However, in two dimensions,
it is possible to eliminate all fermionic correlators thereby obtaining a
closed equation for the Wilson loop (\ref{defloop}).
To this end, let the change of fields under path integral be of a special
type\footnote{We work in the Eucledian
light cone coordinates $x^{\pm}=x^1\pm i x^2$,
denote $\psi=\Bigl({\psi_-\atop \psi_+}\Bigr)$ and use the set of two
dimensional Dirac matrices $\gamma^1=\sigma_1$, $\gamma^2=\sigma_2$.},
\be
\label{transfA}
\arraycolsep 1.5pt
\renewcommand{\arraystretch}{1.5}
\left\{\begin{array}{rcl}\displaystyle
\delta A_+(x)&=&D_+\chi_-(x)=\partial_+\chi_-(x)+i[A_+(x), \chi_-(x)]\\
\delta A_-(x)&=&D_-\chi_+(x)=\partial_-\chi_+(x)+i[A_-(x), \chi_+(x)]\\
\end{array}\right.
\ee
\be
\label{transfB}
\arraycolsep 1.5pt
\renewcommand{\arraystretch}{1.5}
\left\{\begin{array}{rcl}\displaystyle
\delta\psi_+(x)&=&i[\chi_+(x), \psi_+(x)]\\
\delta\psi_-(x)&=&i[\chi_-(x), \psi_-(x)]. \\
\end{array}\right.
\ee
with two arbitrary matrix valued parameters $\chi_+(x)$ and $\chi_-(x)$.
Note that two parameters are exactly what one needs to parametrize
an arbitrary change of vector potential.

Under this transformation the fermionic kinetic term does not change
while the field strength  term does, so that
\be
\label{varact}
\renewcommand{\arraystretch}{2.5}
\begin{array}{rcl}
\delta S[A_{\mu}, \psi]&=&\displaystyle -{1\over g^2}{\rm tr}\int d^2 x
\thinspace[D_+\delta A_(x)--D_-\delta A_+(x)]F_{01}(x)\\
&=&\displaystyle
-{1\over g^2}{\rm tr}\int d^2 x \thinspace[\chi_+(x)-\chi_-(x)]
D_+D_-F_{01}(x).
\end{array}
\ee

However, the transformation (\ref{transfA}, \ref{transfB}) also affects
the path integral measure ${\cal D}\psi$, due to the chiral
anomaly. Using standard methods \cite{Anomaly}\ it is
possible to show that under
(\ref{transfB}) the fermion measure transforms as
\be
\label{anomaly}
{\cal D}\psi_+{\cal D}\psi_-\to{\cal D}\psi_+{\cal D}\psi_-
\exp\left[-{g N\over 4\pi}{\rm tr}\int d^2 x \thinspace
\bigl[\chi_+(x)-\chi_-(x)\bigr]F_{01}(x)
\right].
\ee

Finally, the variation of the path ordered exponential in (\ref{defloop})
yields a contribution
\be
\label{variation}
\arraycolsep 1.5pt
\renewcommand{\arraystretch}{2.5}
\begin{array}{rcl}
&\delta&\displaystyle\left\{{\rm tr}\thinspace
{\rm P}\exp\left[-i\oint  A_{\mu}(y)dy^{\mu}\right]
\right\}\\
&&\displaystyle={\rm tr}\thinspace
{\rm P}\left\{\exp\left[-i\oint  A_{\mu}(y)dy^{\mu}\right]
\left(\oint dx^- D_-\chi_+(x)+\oint dx^+ D_+\chi_-(x)\right)\right\}.\\
\end{array}
\ee
If $\chi_+=\chi_-$ then (\ref{transfA}) is a gauge transformation and
the variation (\ref{variation}) vanishes. That is to say, similarly to
(\ref{varact}, \ref{anomaly}), the right hand side of (\ref{variation})
depends only on the difference $\chi_+-\chi_-$ rather than on $\chi_+$
and $\chi_-$ by themselves.

Demanding that $\delta W[C]/\delta\chi_+(x)=\delta W[C]/\delta\chi_-(x)=0$,
and using the Mandelstam formula \cite{Mandelstam}
\be
\label{mand}
{1\over N}
\left\langle {\rm tr}\left[F_{\mu\nu}(x) {\rm P}
\exp\left(-i\oint  A_{\mu}(y)dy^{\mu}\right)\right]\right\rangle
={\delta W[C]\over \delta \sigma_{\mu\nu}(x)},
\ee
we obtain the loop equation
\be
\arraycolsep 1.5pt
\renewcommand{\arraystretch}{2.5}
\begin{array}{ll}
\label{lpeqn}\displaystyle
\displaystyle\left(\partial_+\partial_--{g^2 N\over 4\pi}\right)&\displaystyle
\left.{\delta W[C]\over
\delta\sigma(x)}\right|_{x=x(\tau)}\\
&\displaystyle =+g^2 \oint dx^-(\tau^{\prime}){\partial\over \partial
x^-(\tau)} \delta^{(2)}(x(\tau)-x(\tau^{\prime}))\left\langle
W_{x(\tau)x(\tau^{\prime})}W_{x(\tau^{\prime})x(\tau)}\right\rangle\\
&\displaystyle =-g^2 \oint dx^+(\tau^{\prime}){\partial\over \partial
x^+(\tau)} \delta^{(2)}(x(\tau)-x(\tau^{\prime}))\left\langle
W_{x(\tau)x(\tau^{\prime})}W_{x(\tau^{\prime})x(\tau)}\right\rangle  . \\
\end{array}
\ee
The right hand side of this equation involves the correlator of Wilson
loops for the two subcontours of $C$ which are obtained by cutting
it at the points $x(\tau)$ and $x(\tau^{\prime})$. Due to the presence of a
delta function it is different from zero
only if $x(\tau)=x(\tau^{\prime})$, so that these subcontours are closed and
$W_{x(\tau)x(\tau^{\prime})}, W_{x(\tau^{\prime})x(\tau)}$
are gauge invariant.
For the same reason as in pure Yang--Mills theory \cite{Kazakov}\ the
contour integrals
in the right hand side of (\ref{lpeqn}) should be understood in the
principal value sense---a small interval of $\tau^{\prime}\in ]\tau-
\epsilon, \tau+\epsilon[$ should be excluded from the integration region.
Then these integrals produce a nonzero contribution only for contours with
self-intersections. A simple, nonselfintersecting Wilson loop obeys,
therefore, the Klein--Gordon equation
\be
\label{kg}
\left(\Delta -{g^2 N\over 4\pi}\right){\delta W[C]\over
\delta\sigma(x)}=0.
\ee
This equation is valid both for finite $N$ and in the large $N$ limit.

An immediate consequence of (\ref{kg}) is that, in contrast to pure
Yang--Mills theory, $W[C]$ can not be merely a function of the total loop
area. If this were the case, $\delta W/\delta\sigma(x)$ would be
an $x$-independent constant which is not a solution of (\ref{kg}).
Instead, as we shall see, (\ref{kg}) has a different solution which for
large contours exhibits a perimeter, rather than the area, law.

To find this solution notice that the expectation value of a Wilson loop in
Yang--Mills theory without fermions can be represented as
\be
W[C]=\exp\left(-{g^2 N\over 2} A\right)=\exp \left[{g^2 N\over 2}
\oint dx^{\mu}dy_{\mu} G(x-y)\right],
\ee
where $G(x-y)$ is the masless propagator defined by $\Delta G(x-y)=
\delta^{(2)}(x-y)$.
Indeed, converting the contour integrals into the area integrals by Stokes'
theorem, we get
$$\oint dx^{\mu}dy_{\mu} G(x-y)=\int d^2x d^2y {\partial\over
\partial x^{\mu}}{\partial\over \partial y^{\mu}} G(x-y)=-\int d^2x d^2 y
\delta^{(2)}(x-y)=-A.$$
Similarly, equation (\ref{kg}) will be satisfied if we consider
\be
\label{gross}
W[C]=\exp \left[{g^2 N\over 2}
\oint dx^{\mu}dy_{\mu} G_m(x-y)\right],
\ee
where $G_m(x-y)$ is now the massive propagator with $m^2=g^2/4\pi$,
satisfying the Klein--Gordon equation $(\Delta -m^2)G(x-y)=
\delta^{(2)}(x-y)$. This fact is easy to check by direct substitution.

For large contours the solution (\ref{gross}) decays like $\propto\exp(-mP)$
where $P$ is the perimeter of the loop. Indeed, the massive propagator
$G_m(x-y)$ vanishes very fast for $|x-y|\gg 1/m$. Thus the contour integral
in (\ref{gross}) is dominated by those $x, y$ which are at most $1/m$
away, giving rise to the perimeter dependence of $W[C]$ for the loops
of large size.

Although (\ref{gross}) satisfies the loop equation exactly, it does not
give the exact expectation value of the Wilson loop in the adjoint
fermion model. The reason is that, unless supplemented by certain boundary
conditions \cite{Migdal}, loop equations may have more than one solution.
However, even without these boundary conditions
it is clear that the area law $W[C]\propto\exp(-\sigma A)$ is
inconsistent with (\ref{kg}), confirming that the Wilson loop is screened in
the massless adjoint model.

\section{Discussion.}

The surprising result of this paper is that certain 1+1 dimensional gauge
theories with massless adjoint fermions exhibit the screening of
fundamental test charges rather than confinement.
Our discussion was focussed on the simplest model, the
$SU(N)/Z_N$ gauge theory with one massless adjoint multiplet.
It is clear, however, that our methods carry over to more complicated
theories, such as those with several massless adjoint multiplets, which
are also in the screening phase.

A somewhat different example, which seems particularly interesting,
is the ${\cal N}=1$ supersymmetric Yang-Mills theory,
\begin{equation}
\sL = \tr\left [i\bar\psi \gamma^\mu D_\mu \psi+
i g\bar\psi\gamma_5[\phi, \psi]+ {1\over 2} D_\mu\phi D^\mu\phi
-{1\over 4 g^2} F_{\mu\nu} F^{\mu\nu} \right ]\ ,
\end{equation}
where $\phi$ is an adjoint scalar and $\psi$ is an adjoint Majorana
fermion. This theory, which may be
obtained by dimensionally reducing the 2+1 dimensional
${\cal N}=1$ SYM theory, was recently discussed in \cite{mats}.
We find that the presence of fermion zero modes in the topologically
non-trivial sectors of $SU(N)/Z_N$ once again guarantees that
$\langle W(C)\rangle=1$ for the contour at infinity.
For $N=2$ it is also possible to show that the model exhibits bilinear
gluino condensation. These results imply that this theory is in
the screening phase. This raises a tantalizing question: is it possible
that $2+1$ and $3+1$ ${\cal N}=1$ supersymmetric Yang-Mills theories
are also in the screening, rather than the confining phase?
We will return to this later.

There are two distinguishing features of 1+1 dimensional gauge theories
which made our analysis possible:
\begin{enumerate}
\item The absence of dynamical degrees of freedom for gauge fields which
leads to trivial Coulomb confinement in pure photodynamics or pure
gluodynamics.
\item The rigid relation of the Wilson loop average to the topological
structure of the theory.
 \end{enumerate}

Neither is true in 3+1 dimensions. This makes even more surprising the
analogy of the phenomenon we observe with the situation in ${\cal N} = 2$
supersymmetric Yang-Mills theory \cite{Seiberg}.
In this theory the confinement of
the unbroken $U(1)$ subgroup sets in as soon as a certain mass term,
which breaks ${\cal N} = 2$ supersymmetry down to ${\cal N} =1$, is added
to the Lagrangian.

What possible lessons could we draw with respect to the
more conventional 4D
theories, in particular to $QCD$ ? The physical case of $QCD$ with
dynamical quarks is well known to
display screening, and we have nothing new to say about
it. The pure 4D Yang-Mills theory is expected to be confining.
In view of what we learned from 1+1 dimensional examples we may wonder,
however, whether instead it could be in the screening phase: certain
collective gluonic excitations might be capable of screening fundamental
test charges. This possibility seems to be experimentally ruled out,
however, since no states of fractional baryon number have been
observed.

A more realistic scenario is
that the pure gluodynamics is confining, while its
${\cal N} =1$ supersymmetric extension is not, due to the presence of
the massless adjoint fermions.
Our 1+1 dimensional examples show that a cloud of
gluinos (with some help from the gluons) can screen a heavy
fundamental charge, and we may be bold enough
to conjecture that this is also possible in 3+1 dimensions.
The screening disappears and the confinement is restored
as soon as the gluinos are given a small mass (and the supersymmetry is
broken). This scenario is sufficiently intriguing that, in our opinion,
it deserves further investigation.

\section*{Acknowledgements}

We are grateful to A. Hashimoto, D. Kutasov, J. Polchinski, M. Shifman,
L. Susskind and A. Vainshtein for interesting discussions.
This work was supported in part by DOE grant DE-FG02-91ER40671,
NSF grant PHY90-21984,
the NSF Presidential Young Investigator Award PHY-9157482, the
James S. McDonnell Foundation grant No. 91-48 and funds
provided  by the U.S. Department of Energy under
cooperative research agreement  DE-FC02-94ER40818.

\end{document}